\documentclass{amsart}
\usepackage{amsfonts,latexsym}
\newcommand{\diff}{\mbox{Diff}_+(S^1)}

\begin{document}

\title[Discrete Lagrangian
systems and Camassa-Holm family]{Discrete Lagrangian
systems on the Virasoro group and Camassa-Holm family}
         \author{A. V. Penskoi}
         \address{Centre de Recherches Math\'ematiques, Universit\'e de
Montr\'eal,
C.P. 6128, Succ. Centre-ville, Montr\'eal, Qu\'ebec, H3C 3J7, Canada}
             \email{penskoi@crm.umontreal.ca}

         \author{A. P. Veselov}
         \address{Department of Mathematical Sciences,
          Loughborough University, Loughborough,
          Leicestershire, LE11 3TU, UK and
         Landau Institute for Theoretical Physics, Moscow, Russia}
         \email{A.P.Veselov@lboro.ac.uk}

\subjclass{34K99, 22E65, 70H99}
\keywords{Camassa-Holm family, Virasoro group, discrete Lagrangian system}

\begin{abstract}
We show that the continuous limit of a wide natural class of the
right-invariant
discrete Lagrangian systems on the Virasoro group gives the family of
integrable
PDE's containing Camassa-Holm, Hunter-Saxton and Korteweg-de Vries equations.
This family has been recently derived by Khesin and Misio\l{}ek
as Euler equations
on the Virasoro algebra for $H^1_{\alpha,\beta}$- metrics. Our result
demonstrates a universal
nature of these equations.
\end{abstract}

\maketitle
\section{Introduction}

Let $M$ be a manifold and $L$ be a function on $M\times M.$
The {\it discrete Lagrangian system} with the Lagrangian $L$
is the system of difference equations
$$ \delta S =0,$$
which describes the
stationary points of the functional
$S=S(X)$ defined on the space of sequences $X=(x_k),x_k \in M,
k\in{\mathbb Z}$ by
a formal sum
$$
S=\sum\limits_{k\in{\mathbb Z}}L(x_k,x_{k+1}).
$$
Sometimes this system has a continuous limit; in that case it is called
a discrete version of the corresponding system of differential equations.

For the integrable equations it is natural to ask for the discretizations,
which are also integrable. The theory of integrable Lagrangian
discretizations of the classical integrable systems
was initiated by J. Moser and one of the authors in \cite{V1, MV1}.

In particular, it was shown that the discrete Lagrangian system
on the orthogonal group $O(n)$ with the Lagrangian
$L(X,Y) =tr( XJY^{T}),$ $X,Y\in O(n), J=J^{T}$ can be considered
as an integrable discrete version of the Euler-Arnold top \cite{Arn}.

The first attempt to generalize this approach to the infinite-dimensional
situation was done in \cite{MV2}, where the case of the group of
area-preserving
plane diffeomorphisms $\mbox{SDiff}({\mathbb R}^2)$ was considered.
This was followed by the papers \cite{P1}, \cite{P2}, \cite{P3} where
the discrete Lagrangian systems
on the Virasoro group have been discussed. The interest to
the Virasoro group was motivated by the important observation due to
Khesin and Ovsienko~\cite{KO} that the Korteweg-de~Vries
equation can be interpreted as an Euler equation on the Virasoro algebra.
The goal was to find an integrable Lagrangian discretization of the
KdV equation.
The question is still open but some good candidates for the answer
have been found (see~\cite{P2}).

In this note inspired by \cite{P3} and the recent, very interesting
Khesin-Misio\l{}ek paper \cite{KM}
we are looking at the same problem but from a different angle. We
consider a wide natural class of the
discrete Lagrangian systems on the Virasoro
group and look what we can get in the continuous limit.

The result turns out to be surprising:
in spite of the fact that the class of discrete systems we consider
is quite general in the continuous limit
we have the following family of integrable (!) equations:
\begin{equation}
\label{CH}
\alpha(v_t + 3vv_x) - \beta(v_{xxt} + 2v_xv_{xx} +  v v_{xxx}) - bv_{xxx} = 0,
\end{equation}
$\alpha, \beta$ and $b$ are arbitrary constant parameters.

This family appeared in Khesin and Misio\l{}ek paper \cite{KM} as the
Euler equations
on the Virasoro algebra for a natural two-parameter family of metrics.
We will call it {\it Camassa-Holm family} because for the non-zero
$\alpha$ and $\beta$
(i.e. in generic case) (\ref{CH}) is equivalent to the Camassa-Holm equation
\cite{CH}.\footnote{We should warn the reader that there is another
family containing Camassa-Holm equation discussed recently by
Degasperis, Holm and Hone in \cite{DHH}.
Their family besides Camassa-Holm equation contains only one more
integrable case,
so in general is non-integrable.} In the degenerate cases we have the
Hunter-Saxton equation \cite{HS}
corresponding to $\alpha=0, \beta \neq 0$
and the KdV equation (when $\beta =0, \alpha \neq 0, b \neq 0).$
As it was shown in \cite{KM} these three cases correspond precisely
to the three
different types of generic Virasoro orbits.

\section{Discrete Lagrangian systems on the Virasoro group and their
continuous limit}

Let  $\diff$ be the group of orientation preserving diffeomorphisms of $S^1$.
We will represent an element of $\diff$ as a function
$f:{\mathbb R}\rightarrow{\mathbb R}$ such that
\begin{enumerate}
\item $f\in C^\infty({\mathbb R}),$
\item $f'(x)>0,$
\item $f(x+2\pi)=f(x)+2\pi.$
\end{enumerate}
Of course, such representation is not unique: the functions $f$ and $f+2\pi$
represent the same element of $\diff.$

This group has nontrivial central extension defined by
the so-called Bott cocycle, which is unique up to an isomorphism.
This extension is called the {\it Virasoro group} (also known as {\it
Bott-Virasoro group}) and is denoted as
$\mbox{Vir}.$ Elements of $\mbox{Vir}$ are pairs $(f,F)$,
where $f\in\diff,$
$F\in{\mathbb R}.$
The product of two elements in $\mbox{Vir}$ is defined as
$$
(f,F)\circ(g,G)=(f\circ g,F+G+\int\limits_{0}^{2\pi}\log(f\circ g)'\,d\log g').
$$

The unit element $e$ of $\mbox{Vir}$ is $(id,0).$ The inverse element of
$(f,F)$ is
$(f^{-1},-F).$

Let us describe the class of discrete Lagrangian systems on $\mbox{Vir}$ we
are going to consider.

The main property we impose is the right-invariance of the Lagrangian:
\begin{equation}\label{Inv}
L(X,Y) = L(Xg, Yg),\quad X,Y,g \in \mbox{Vir}.
\end{equation}
Because of this property one can rewrite
the Lagrangian in the form
\begin{equation}\label{L}
L(X,Y) = L(X Y^{-1}, e) = H(X Y^{-1}),
\end{equation}
where $H(X) = L(X,e).$
Thus any right-invariant discrete Lagrangian $L$ is determined by the
corresponding function $H$ on this group.

We will consider the discrete Lagrangian systems on $\mbox{Vir}$ corresponding
to the functions $H$ of the following form
\begin{equation}\label{finalH}
H((f,F))=F^2+\int\limits_0^{2\pi}V(f(x)-x,f'(x))\,dx,
\end{equation}
where $f$ is a diffeomorphism, $F\in\mathbb{R},$ and $V(x_1,x_2)$ is an
arbitrary, $2\pi$-periodic in $x_1$
function of two variables, which satisfies the condition:
$$V_1(0,1) =0.$$
Here and below the function with the indices $1,$ $2$ or $3$
stands for the corresponding partial derivatives:
$V_1 = \frac{\partial V}{\partial x_1},
V_{11} = \frac{\partial^2 V}{\partial x_1^2}$ etc.

Periodicity of $V$ is related to the fact that $f(x)$ and $f(x)+2\pi$
represent the same diffeomorphism of $S^1.$
The difference $f(x)-x,$ which is a $2\pi$-periodic function, is
as natural as $f(x)$ itself, so the only property which might look artificial
is the last condition on the partial derivative of $V$.
In order to explain its role and the choice of the Lagrangians (\ref{finalH})
let us consider more general functionals:
\begin{equation}\label{functionH}
H((f,F))=F^2+\int\limits_0^{2\pi}U(f(x),f'(x),x)\,dx,
\end{equation}
where $U(x_1,x_2,x_3)$ is an arbitrary function $2\pi$-periodic with respect to
the first argument.
Thus, we consider the functional
$$
S=\sum\limits_{k\in{\mathbb Z}}L((f_k,F_k),(f_{k+1},F_{k+1})),
$$
where $\{(f_k,F_k)\}$ is a sequence of points on $\mbox{Vir}$
and $L$ is defined
using the function $H$~(\ref{functionH}) as described above:
$$
L((f_k,F_k),(f_{k+1},F_{k+1}))=H((f_k,F_k)\circ(f_{k+1},F_{k+1})^{-1}).
$$
The discrete Euler-Lagrange equations for this functional
have the form
\begin{equation}\label{EL1}
-\Omega_k+\Omega_{k+1}=0,
\end{equation}
and
$$
-2\Omega_k\left[\log(\omega'_k)\right]''-U_1(\omega_k,\omega'_k,x)\omega'_k+%
U_{12}(\omega_k,\omega'_k,x)(\omega'_k)^2+%
U_{22}(\omega_k,\omega'_k,x)\omega''_k\omega'_k+
$$
$$
+U_{23}(\omega_k,\omega'_k,x)\omega'_k+%
2\Omega_{k+1}\left[\log((\omega^{-1}_{k+1})')\right]''+%
U_1(x,\frac{1}{(\omega^{-1}_{k+1})'},\omega^{-1}_{k+1})(\omega^{-1}_{k+1})'-
$$
$$
-U_{12}(x,\frac{1}{(\omega^{-1}_{k+1})'},\omega^{-1}_{k+1})+%
U_{22}(x,\frac{1}{(\omega^{-1}_{k+1})'},\omega^{-1}_{k+1})%
\frac{(\omega^{-1}_{k+1})''}{((\omega^{-1}_{k+1})')^2}-
$$
\begin{equation}\label{EL2}
-U_{23}(x,\frac{1}{(\omega^{-1}_{k+1})'},\omega^{-1}_{k+1})%
(\omega^{-1}_{k+1})'=0,
\end{equation}
where $(\omega_k,\Omega_k)$
are discrete analogues of angular velocities:
$$
(\omega_k,\Omega_k)=(f_{k-1},F_{k-1})\circ(f_k,F_k)^{-1},\quad k\in\mathbb{Z}.
$$

The first equation~(\ref{EL1}) simply says that
$\Omega_{k+1}=\Omega_k,$ so $\Omega_k$ is an integral of our discrete
Lagrangian system.

Let us now find a continuous limit of our Euler-Lagrange equations.
To do this we suppose that the angular velocity is of the form
$$
(\omega_l,\Omega_l)=(id+\varepsilon v_l(x),\varepsilon A_l),
$$
i.~e. the angular velocity is the identity element of
$\mbox{Vir}$ up to $O(\varepsilon).$
Also we assume that
\begin{equation}\label{cl}
v_k(x)=v(x,t),\quad A_k=A(t),\quad v_{k+1}(x)=v(x,t+\varepsilon),\quad%
A_{k+1}=A(t+\varepsilon).
\end{equation}
We should substitute these formulae in the discrete Euler-Lagrange equations
and take the term of lowest order with respect
to $\varepsilon$ in the corresponding Taylor series. This is what we mean by
the continuous limit of our discrete system.

It is easy to see that the continuous limit ot the
first Euler-Lagrange equation~(\ref{EL1})
is $A_t=0,$ i.e. $A(t) =A$ is a constant.
But when we consider the continuous limit of the second
Euler-Lagrange equation~(\ref{EL2}) we have the following problem: the
leading $\varepsilon^1$-term
gives us an ordinary (but not a partial) differential equation.
Indeed a straightforward
calculation shows that
this term has a form
\begin{equation}\label{1term}
U_{233}(x,1,x)v-2U_1(x,1,x)v_x+2U_{23}(x,1,x)v_x+2U_{123}(x,1,x)v+
\end{equation}
$$
+U_{112}(x,1,x)v-U_{11}(x,1,x)v+2U_{12}(x,1,x)v_x-U_{13}(x,1,x)v.
$$

Since we want to have a more interesting continuous limit in $x$, we have
to eliminate this term and look at the next $\varepsilon^2$-term.
The condition that $\varepsilon^1$-term is equal to zero is equivalent
to some relations for partial
derivatives of $U$ in the points $(x,1,x).$ It is not clear how to
resolve these
equations in a general case, but one can easily check that there is one
important particular class of solutions:
$$
U(x_1,x_2,x_3)=V(x_1-x_3,x_2),
$$
where $V(x_1,x_2)$ is an arbitrary $2\pi$-periodic in $x_1$
function with the only condition:
$$V_1(0,1) = 0.$$

Thus we arrive at the Lagrangians of the form (\ref{L}),(\ref{finalH}).
For them (\ref{1term}) is identically equal to zero and after a very long
but straightforward calculation we end up with the
second order term, which gives
$$
V_{11}(0,1)(v_t-3vv_x)-V_{22}(0,1)(v_{xxt}-2v_{xx} v_x-v
v_{xxx})-4Av_{xxx}=0.
$$
If we change $t\mapsto -t$ and introduce
$$\alpha=V_{11}(0,1),\beta=V_{22}(0,1),b=-4A,$$ we come to the
Camassa-Holm
family
as it appears in Khesin-Misio\l{}ek paper (eq. 3.7 in \cite{KM}):
\begin{equation}
\label{CH1}
\alpha(v_t + 3vv_x) - \beta(v_{xxt} + 2v_xv_{xx} +  v v_{xxx}) - bv_{xxx} = 0.
\end{equation}
This is the continuous limit of the discrete systems with the
Lagrangians (\ref{L}), (\ref{finalH}).

Formally this family has three parameters, but we have a freedom of
multiplication of the equation
by a non-zero constant, the two-dimensional scaling symmetry group
$v \rightarrow \lambda v, t \rightarrow \mu t, x \rightarrow \lambda \mu x$
and the Galilean group $v \rightarrow v + c, x \rightarrow x + dt, t
\rightarrow t.$
Modulo these symmetries we have just one generic orbit, containing
the equation with
$\alpha = 1, \beta = 1, b =0:$
$$
v_t - v_{xxt} + 3v v_x - 2v_x v_{xx} - v v_{xxx}  = 0,
$$
which is one of the canonical forms of the {\it Camassa-Holm
shallow-water equation} \cite{CH}
$$
v_t + 2\kappa v_x + \gamma v_{xxx} - v_{xxt}  + 3v v_x - 2v_x v_{xx}
- v v_{xxx}  = 0.
$$

We have also four degenerate orbits.
When $\alpha \neq 0, \beta = 0, b \neq 0$
the equation (\ref{CH1}) is equivalent to the KdV equation:
$$v_t + 3 v v_x + v_{xxx} = 0.$$
Further degeneration $\alpha \neq 0, \beta = 0, b = 0$ leads to the
dispersionless KdV equation (sometimes called also Hopf equation):
$$v_t + 3 v v_x = 0.$$
When $\alpha = 0, \beta \neq 0$ we have the {\it Hunter-Saxton
equation} \cite{HS}
$$v_{xxt} + 2v_xv_{xx} +  v v_{xxx} =0.$$
Finally if both $\alpha$ and $\beta$ are zero (but $b$ is not) we simply have
$$v_{xxx}=0.$$

\section{Discussion}

The fact that the continuous limit of a wide natural class of the
discrete right-invariant
Lagrangian systems on the Virasoro group is integrable seems to be remarkable.
This universality of the Camassa-Holm family is closely related with
the results \cite{KM},\cite{M} by Khesin and Misio\l{}ek
who interpreted this family as
the Euler equtions on the Virasoro algebra corresponding to the special
family of the right-invariant metrics on the Virasoro group
($H^1_{\alpha,\beta}$- metrics).
The heuristic explanation of our result is that these equations
can be considered as nonlinear analogues of the harmonic oscillators
on the Virasoro group: in the first approximation all Hamiltonian
systems near equilibriums behave like harmonic oscillators.

The integrability of the Camassa-Holm family can be shown
in different ways (see e.g. \cite{KM}, \cite{BSS}) but the underlying
reasons for that seem to be deep and deserve better understanding.
In particular, it would be interesting to consider the discrete
right-invariant Lagrangians depending
on the derivatives of a diffeomorphism $f$ up to the second order to see
if similar phenomenon holds in that case or not.
Another interesting question is what happens for other infinite-dimensional
groups (e.g for $\mbox{SDiff}({\mathbb R}^2)$).

It is instructive to compare the situation with the finite-dimensional case.
For example, for the orthogonal group $O(n)$
we may have as a continuous limit of a right-invariant discrete
Lagrangian system any right-invariant geodesic flow, which is known
for $n > 3$ to be
in general  non-integrable (see e.g. \cite{V2}). There are integrable cases
of the Euler equations on $O(n)$ (for example the Manakov metrics
\cite{Manakov}) but no analogues of our result are known for them.

It is interesting to mention that from Khesin-Misio\l{}ek results
(see section 5 in \cite{KM})
it follows that $H^1_{\alpha,\beta}$-metrics can be considered as the
analogues of the very special cases of the
Manakov metrics. The question what are the analogues for
the general Manakov metrics (if there are any) seems to be open.

\section*{Acknowledgments}
We are grateful to Boris Khesin and Andy Hone for useful
and stimulating discussions. We would like also to thank a referee
for useful comments.

One of the authors (A. P.) is very grateful to the Centre de Recherches
Math\'ematiques (CRM) for its hospitality.

\end{document}